\documentclass[aps, reprint, showpacs, epsfig]{revtex4-1}
\usepackage{graphicx}
\usepackage{subfigure}
\usepackage{epsfig}
\usepackage{latexsym}
\usepackage{epstopdf} 
\usepackage{color}
\usepackage{amsmath}
\usepackage{mathtools}

\usepackage{hyphenat}
\let\oldref\ref
\renewcommand{\ref}[1]{(\oldref{#1})}
\usepackage[colorlinks=true,linkcolor=blue,urlcolor=blue,citecolor=blue]{hyperref}
\usepackage{amsmath}

\begin{document}
\title{Understanding the phenomenon of viscous slowing down of glass-forming liquids from the
static pair correlation function}
\author{Ankit Singh and Yashwant Singh}
\affiliation{Department of Physics, Banaras Hindu University,
Varanasi-221 005, India.}
\date{\today}

\begin{abstract}
A theory which uses data of the static pair-correlation function is developed to calculate quantities 
associated with the viscous slowing down of supercooled liquids. We calculate value of the energy 
fluctuations that determine the number of ``stable bonds'' a particle forms with neighbors from data
of the structural relaxation time. The number of bonds and the activation energy for relaxation are 
shown to increase sharply in a narrow temperature range close to the glass temperature.
The configurational entropy calculated from values of the ``configurational fluctuations'' is found 
in good agreement with the value determined from simulations.
\end{abstract}


\maketitle

When a liquid is supercooled bypassing its crystallization, it continues to remain structurally disordered,
but its dynamics slows down so quickly that below a temperature, called the glass temperature $T_{g}$,
the structural relaxation takes such a long time that it becomes difficult to observe \cite{Angell}.
Below $T_{g}$  the liquid appears to be trapped virtually for ever in one of many possible amorphous 
structures. The ubiquity of the phenomenon, irrespective of molecular details points to a collective or 
cooperative behavior characterized by a length scale that sharply grows close to the glass temperature.
Since the characteristic features of the static structure factor remain unchanged when the system is cooled,
claim was made \cite{Cavagna} that it is impossible to use the static correlation function to explain the 
sharp rise in the relaxation time. This observation led many to look for a ``hidden length'' \cite{Ediger} 
or an ``amorphous'' \cite{Montanari,Sausset} or ``frustrated'' 
\cite{GTarjus} order in liquids to which the pair correlation function is essentially blind. However, contrary to 
the widely held view that the static correlation function cannot explain the slowing down of dynamics, 
we show that it, indeed, provides information that can be used to 
understand and calculate quantities associated with the glass formation.

A particle in a liquid has the kinetic energy which gives it motion and the effective potential energy 
due to its interactions with neighboring particles who restricts its motion by creating a cage with 
barrier. The competition between the kinetic energy which has a Maxwellian distribution and the 
effective potential energy in a supercooled liquid can create a very mosaic situation in respect
of particles motion and distribution . When height of the barrier becomes larger than the total 
energy of a particle, it gets trapped and localized in the cage. On the other hand , a particle 
which total energy is higher than the barrier moves around and collides with other particles.
The concentration of these particles depends on density and temperature. A deeply supercooled liquid
can be considered as a network of trapped particles connected with each other by (non-chemical) bonds
and few free particles which number decreases on decreasing the temperature. The life-time of bonds
which may vary from microscopic to  macroscopic time due to energy fluctuations, depends on their bonding 
energies. Close to $T_{g}$, both the number of bonds and the bonding energy with which a particle
is bonded with neighbors are likely to increase sharply and the structural relaxation becomes a 
thermally activated process over the barrier height. We calculate the barrier height (or activation
energy) from the data of the radial distribution function \cite{Singh}.

The system we consider is the 
Kob-Anderson 80:20 mixture of Lennard-Jones particles consisting of two species of particles a and b
\cite{Kob}. All particles have the same mass m and the interaction between two particles of type 
$\alpha$,$\gamma$ $\in$ $[a,b]$ is given by
\begin{equation}
 u_{\alpha\gamma}(r)= 4\epsilon_{\alpha\gamma} 
[(\frac{\sigma_{\alpha\gamma}}{r})^{12}-(\frac{\sigma_{\alpha\gamma}}{r})^{6}] ,
\end{equation}
with $\epsilon_{aa}=1$, $\sigma_{aa}=1$, $\epsilon_{ab}=1.5$, $\sigma_{ab}=0.8$, $\epsilon_{bb}=0.5$, 
$\sigma_{bb}=0.88$. Length, energy and temperature are given in units of $\sigma_{aa}$, $\epsilon_{aa}$ 
and $\epsilon_{aa}/k_{B}$, respectively.
Values of the radial distribution function used in the present calculation for density 
$\rho=1.20$ and temperature range $T$ $\in$ $[0.45,1.00]$ were evaluated using molecular dynamics 
simulation by Das {\it et al.} \cite{Tah}.

The radial distribution function which for a simple liquid is defined as \cite{Hansen}
\begin{equation}
g(\lvert\vec{r_{2}}-\vec{r_{1}}\rvert) \equiv g(r) = 
\frac{1}{N\rho} \langle \sum_{j}^{N}\sum_{j \neq k}^{N} 
\delta(\vec{r}-\vec{r_{j}}+\vec{r_{k}}) \rangle ,          
\end{equation}
where $N$ is number of particles, $\rho$, the number density and the angular bracket represents the 
ensemble average, tells us what is probability of finding a particle at a distance $r$ from a 
reference (central) particle. The average number of particles lying within the range $r$ and
$r+\mathrm{d}r$ from the central particle is $4\pi\rho g(r)r^{2}\mathrm{d}r$.
As $g(r)$ defined by Eq. (2) has no information about the kinetic energy of particles,
one cannot say how many of these particles are trapped in the cage and how many are free.
To find this we rewrite $g_{\alpha\gamma}(r)$ in the center-of-mass coordinates as \cite{Singh}, 

\begin{equation}
 g_{\alpha\gamma}(r)=\left(\frac{\beta}{2\pi\mu}\right)^{\frac{3}{2}}\int \mathrm{d}{\bf p} \  
\mathrm{e}^{-\beta (\frac{p^2}{2\mu} + w_{\alpha\gamma}(r))} , 
\end{equation}
where $\beta$ is the inverse temperature measured in units of 
the Boltzmann constant $k_{B}$ which we take henceforth as unity, 
${\bf p}$  is the relative momentum of a particle of mass $\mu=m/2$. 
The effective potential $w_{\alpha\gamma}(r)=-T\ln g_{\alpha\gamma}(r)$ 
is sum of the (bare) potential and the system-induced potential 
energy of interaction between a pair of particles of species $\alpha$ and $\gamma$
separated by distance $r$ \cite{Hansen}.
The peaks and troughs of $g_{\alpha\gamma}(r)$ create minima and maxima in $\beta w_{\alpha\gamma}(r)$
as shown in Fig. 1 for potential of Eq. (1) at $\rho=1.20$ and $T=0.45$.
We denote a region between two maxima $i-1$ and $i$ $(i\geq 1)$ as $i$th shell and the minimum
of the shell by $\beta w_{\alpha\gamma}^{(i,d)}$.
The value of $i$th maximum is denoted by $\beta w_{\alpha\gamma}^{(i,u)}$ and its location by $r_{ih}$.
\begin{figure}[t]
\includegraphics[width=0.4\textwidth]{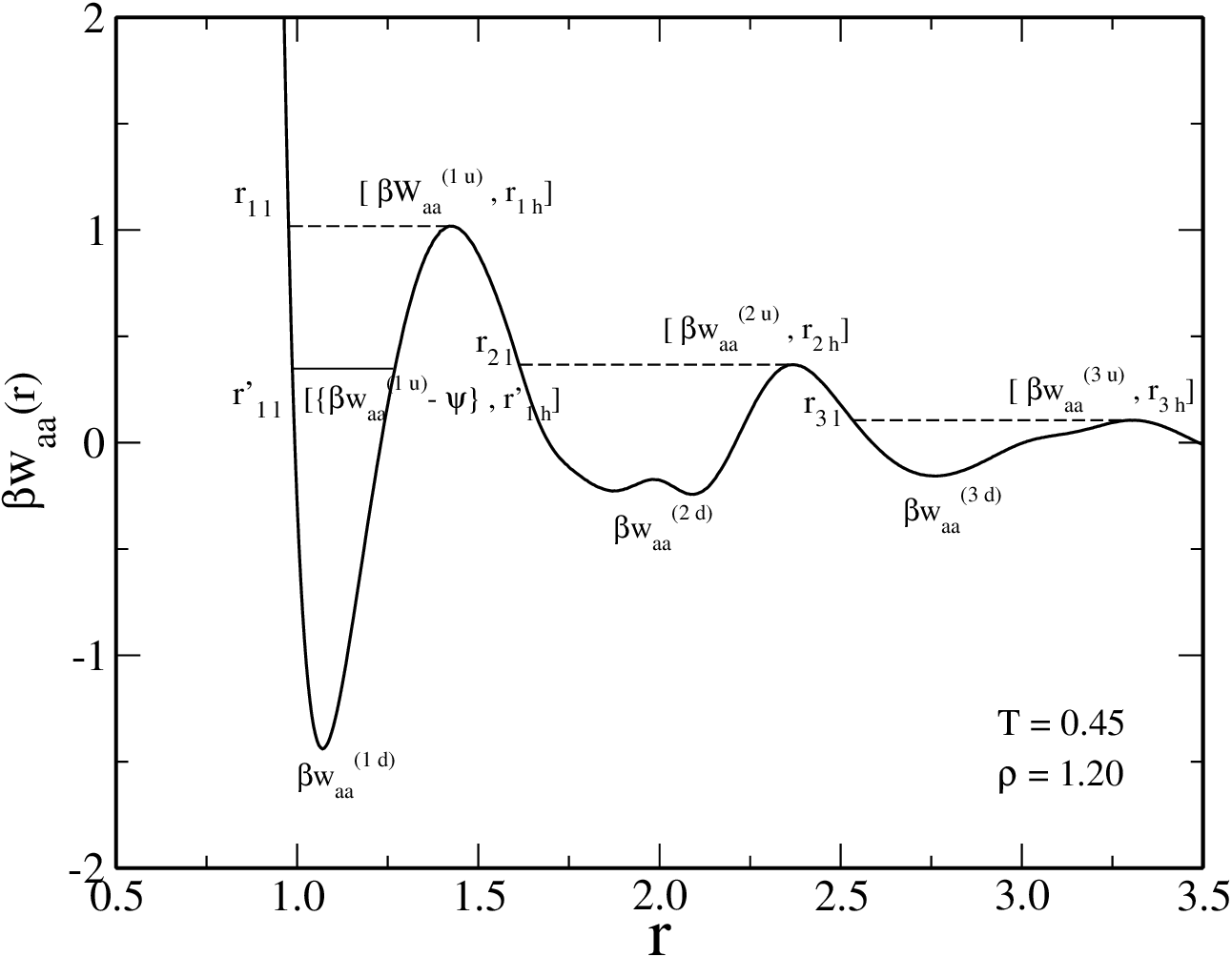}
\caption{The reduced effective potential $\beta w_{aa}(r)$ between a pair of particles of species $\alpha$
and $\gamma$ separated
by distance r (expressed in unit of $\sigma_{aa}$) in a system of Lennard-Jones at a
density $\rho=1.20$ and temperature $T=0.45$. $\beta w_{aa}^{(iu)}, r_{ih}$ are, respectively,
value and location of $i$th maximum and $r_{il}$ is the location on the left hand side of the shell where
$\beta w_{aa}^{(i)}(r)=\beta w_{aa}^{(iu)}$ (shown by dashed line).
The location $r'_{il}$ and $r'_{ih}$ are values of r on the left and the right hand side of the shell
where $\beta w_{aa}^{(i)}(r)=[\beta w_{aa}^{(iu)}-\psi]$ (shown by full line). $\beta w_{aa}^{(id)}$ is
the depth of the $i$th shell.}
\label{fig-1}
\end{figure}

All those particles of $i$th shell which energies are less or equal to $\beta w_{\alpha\gamma}^{(i,u)}$
i.e. $\beta[\frac{p^2}{2\mu}+w_{\alpha\gamma}^{(i)}(r)] \leq \beta w_{\alpha\gamma}^{(i,u)}$ 
will get trapped in the shell and can be considered to be bonded with the central particle.
The number of bonded particles is found from a part of $g_{\alpha\gamma}(r)$ defined as
%\begin{equation}
\begin{eqnarray}\nonumber
g_{\alpha\gamma}^{(ib)}(r) & =& 4\pi(\frac{\beta}{2\pi\mu})^{3/2} \mathrm{e}^{-\beta w_{\alpha\gamma}^{(i)}(r)} 
\int_{0}^{\sqrt{2\mu[w_{\alpha\gamma}^{(iu)}-w_{\alpha\gamma}^{(i)}(r)]}}      \\
&&  \ \times \mathrm{e}^{-\beta p^2/2\mu} p^2 \mathrm{d}p ,
\end{eqnarray}
%\end{equation}
where $w_{\alpha\gamma}^{(i)}(r)$ is the effective potential in the range $r_{il}\leq r \leq r_{ih}$
of $i$th shell. Here $r_{il}$ is the value of $r$ where $w_{\alpha\gamma}^{(i)}(r)= w_{\alpha\gamma}^{(iu)}$
on the left hand side of the shell (see Fig. 1). The number of particles of $i$th shell bonded with the
central particle of $\alpha$ species is
\begin{equation}
n_{\alpha}^{(ib)} = 4\pi \sum_{\gamma}\rho_{\gamma}\int_{r_{il}}^{r_{ih}} 
g_{\alpha\gamma}^{(ib)}(r) r^2 \mathrm{d}r  ,
\end{equation}
where $\rho_{\gamma}$ is density of $\gamma$ species. The total number $N_{\alpha}^{(b)}$ is found from
$N_{\alpha}^{(b)}= \sum_{i} n_{\alpha}^{(ib)}$.
This number is found to increase rapidly as temperature is lowered (shown in Fig. 4).

It may, however, be noted that the energy fluctuations present in the system 
will make many of the bonded particles to overcome the barrier and become free. The energy fluctuation 
$(\Delta E^{2} = \langle E^{2} \rangle - \langle E \rangle^{2})$ is expressed in terms of the constant
volume specific heat \cite{Reif}
\begin{equation}
\sqrt{\dfrac{\Delta E^{2}}{N}} = \sqrt{C_{v}(T)}\, \ T ,
\end{equation}
where $C_{v}$ is the specific heat per particle. Its value is calculated from $g_{\alpha\gamma}(r)$ 
\cite{Barrat}. The entropy of the system is 
related with $C_{v}$ as
\begin{equation}
T \dfrac{\mathrm{d}S(T)}{\mathrm{d}T} = C_{v}(T) .
\end{equation}

\begin{figure}[t]
\includegraphics[width=0.47\textwidth]{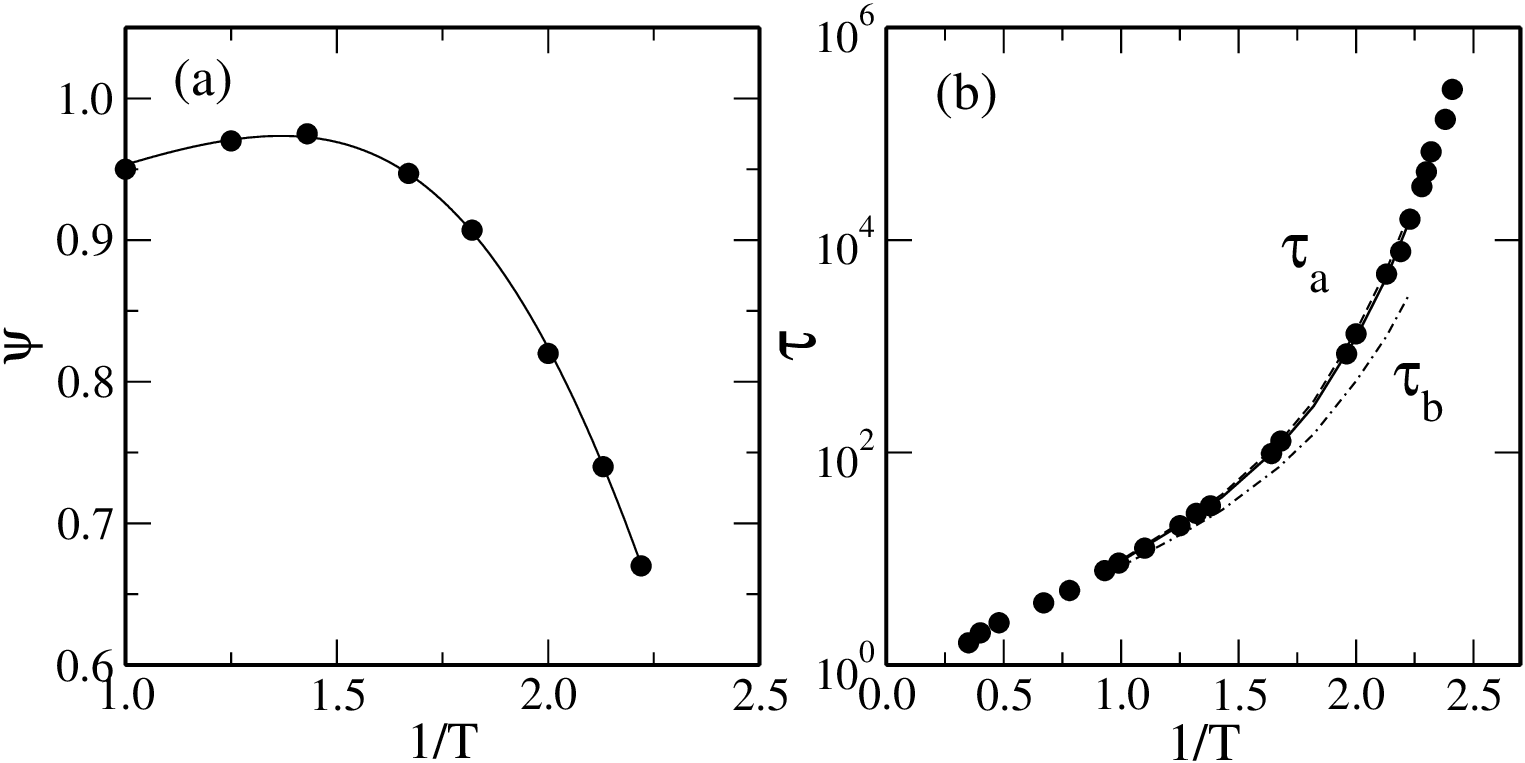}
\caption{{(a) Plot of $\psi(T)$ vs $\frac{1}{T}$. Circles represent the calculated values and the solid line
values calculated from Eq. (11).}
{(b) Comparison of calculated values of $\tau$ (solid line) with the values given in Ref. \cite{Tarjus}
(filled circles). Values of $\tau_{a}$ (dashed line) and $\tau_{b}$ (dash-dotted line) are plotted to
show their relative contributions.}}

\label{fig-2}
\end{figure}

The configurational entropy is defined as $S_{conf}(T) = S(T) - S_{g}(T)$ where $S_{g}(T)$
is the entropy of an equilibrium system which is trapped in a particular amorphous configuration (glass) 
\cite{Berthier}.

The assumption we make here is that the fluctuations that give rise the configurational entropy are also 
responsible for stabilizing the number of bonded particles.
We call this part of fluctuations as the configurational fluctuations and denote as $\psi(T) {T}$.
The specific heat of the glass, $C_{vg}$ is found by subtracting $\psi^{2}$ from $C_{v}$, {\it i.e.},
$C_{vg}=C_{v}-\psi^{2}$. One can find values of $\psi(T)$ from above equations as both $C_{v}(T)$ and 
$C_{vg}(T)$ (or $S(T)$ and $S_{g}(T)$) can be calculated from the inter-particle interactions \cite{Berthier}.
However, instead of finding value of $\psi(T)$ in this way, we take it as an adjustable parameter 
and determine its value from known value of the structural relaxation time $\tau$ following a method
described below and then use it to calculate $S_{conf}(T)$.

The configurational fluctuations will make the life-time of all those bonded particles of $i$th shell 
whose energies lie between $\beta w_{\alpha\gamma}^{(iu)}$ and $[\beta w_{\alpha\gamma}^{(iu)} - \psi]$ 
so short that their role in creating the cage is negligible. We call all these particles as 
metastably bonded particles (henceforth referred to as $m$ particles) and all those particles whose 
energies lie between $[\beta w_{\alpha\gamma}^{(iu)} - \psi]$ and $\beta w_{\alpha\gamma}^{(id)}$ as 
stably bonded particles (henceforth referred to as $s$ particles).
The number of $s$-particles can be found from Eqs. (4) and (5) by replacing $g_{\alpha\gamma}^{(ib)}(r)$
in Eq. (5) by $g_{\alpha\gamma}^{(is)}(r)$ which in turn is found from Eq. (4) by changing the upper limit of
integration from ${\sqrt{2\mu[w_{\alpha\gamma}^{(iu)}-w_{\alpha\gamma}^{(i)}(r)]}}$ to
${\sqrt{2\mu[w_{\alpha\gamma}^{(iu)}-\psi T- w_{\alpha\gamma}^{(i)}(r)]}}$ \cite{Singh}. The number of metastably bonded
particles $N_{\alpha}^{(m)}$ is equal to the number of bonded particles $N_{\alpha}^{(b)}$ minus the number of 
$s$-particles $N_{\alpha}^{(s)}$.
\begin{figure}[t]
\includegraphics[width=0.4\textwidth]{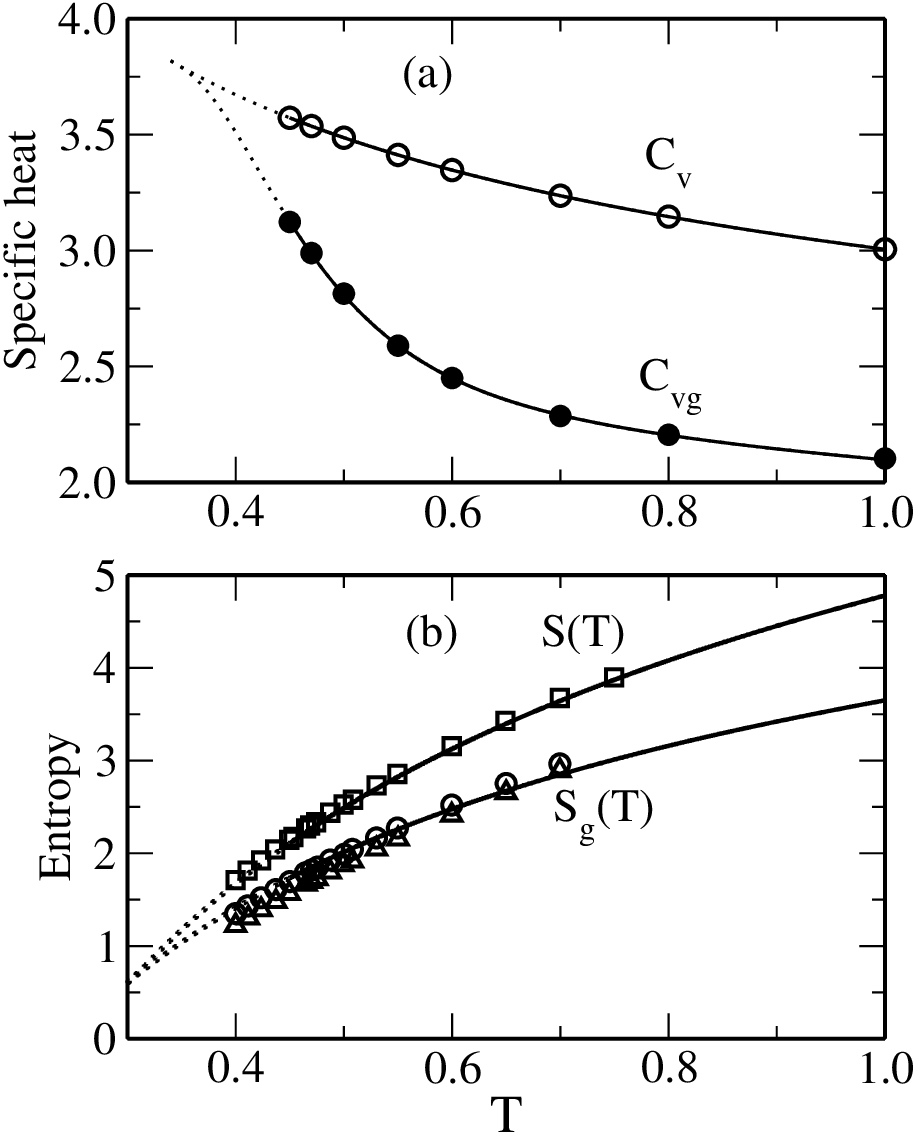}
\caption{{(a) Comparison of constant volume specific heat $C_{v}$ of the system with that of the glass
$C_{vg}=C_{v}-\psi^{2}$ as a function of T. The circles represent the calculated values and the solid
lines represent values found from the algebraic fit as discussed in the text.}
{(b) Comparison of calculated values of $S(T)$ and $S_{g}(T)$ (solid lines) with values found from
simulations \cite{Berthier} (squares represent $S(T)$ and circles (harmonic) and triangles (anharmonic)
represent $S_{g}(T)$).}}

\label{fig-3}
\end{figure}

The activation energy, of relaxation is equal to the energy with which a particle is bonded with
$s$-particles. Thus the activation energy for a particle of species $\alpha$ is 

%\begin{equation}
\begin{eqnarray}\nonumber
\beta E_{\alpha}^{(s)}(T) &=& 4\pi \sum_{\gamma}\rho_{\gamma} \sum_{i} \int_{r'_{il}}^{r'_{ih}} 
[\beta w_{\alpha\gamma}^{(iu)}-\psi- \beta w_{\alpha\gamma}^{(i)}(r)] \\
&& \ \times g_{\alpha\gamma}^{(is)}(r) r^2 \mathrm{d}r .
\end{eqnarray}
%\end{equation}

The energy is measured from the effective barrier height $[\beta w_{\alpha\gamma}^{(iu)}-\psi]$.
The relaxation time of species $\alpha$ is 
\begin{equation}
\tau_{\alpha}(T) = \tau_{0} \exp{[\beta E_{\alpha}^{(s)} (T)]} ,
\end{equation}
where $\tau_{0}$ is a microscopic time scale. The observed relaxation time $\tau$ for a system of binary
mixture is
\begin{equation}
\tau = x_{a} \tau_{a} + x_{b} \tau_{b} ,
\end{equation}
where $x_{\alpha}$ is the concentration of species $\alpha$.

Following the procedure described above we determined values of $\psi(T)$ from the data of $\tau$
reported by Berthier and Tarjus \cite{Tarjus}. In Fig. 2(a) we plot $\psi(T)$ vs $\frac{1}{T}$ and in Fig. 2(b)
calculated values of $\tau$ are compared with those given in Ref. \cite{Tarjus}. In Fig. 2(b) we also plot values of 
$\tau_{a}$ and $\tau_{b}$ to show their relative contributions. From Fig. 2(a) one notes that 
$\psi(T)$ is maximum at $T\simeq 0.7$ and declines sharply on lowering the temperature while on the higher 
temperature side its decline is only marginal. The decline found for $T \gtrsim 0.70$ is possibly due to 
emergence of additional mechanism such as the one described by the mode coupling theory \cite{Gotze} in
addition to the activation for the relaxation. The decline in values of $\psi(T)$ for $T\lesssim0.7$ measures 
the decreasing influence of fluctuations on bonded particles.
As $\psi(T)$ decreases more and more bonded particles become $s$-particles which in turn increase the 
activation energy. As argued below this increase may be very sharp in a narrow temperature range.
\begin{figure}[]
\includegraphics[width=0.4\textwidth]{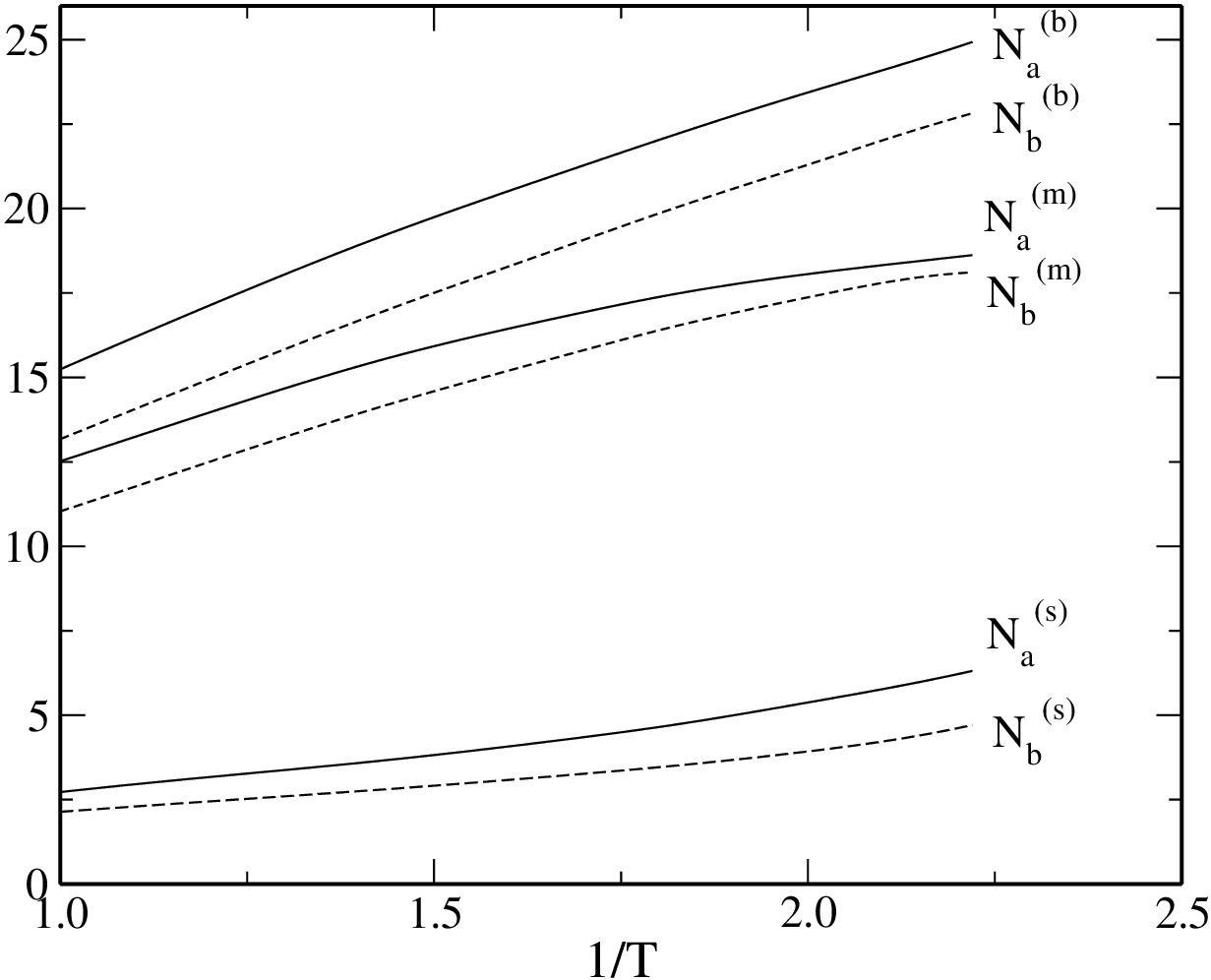}
\caption{Values of $N_{a}^{(b)}$, $N_{a}^{(m)}$, $N_{a}^{(s)}$ (solid lines) and
$N_{b}^{(b)}$, $N_{b}^{(m)}$, $N_{b}^{(s)}$ (dashed lines) are plotted as a function of $\frac{1}{T}$.}

\label{fig-4}
\end{figure}

To fit the data of $\psi(T)$ in the temperature range $T$ $\in$ $[0.45,1.00]$ and to extrapolate to
lower temperatures we use a functional form,
\begin{equation}
\psi(T) = a_{0} \sqrt{C_{v}}\, \ [1-\exp{\{-b_{0} \frac{(T-T_{c})}{T_{c}}\}}]^{\delta} ,
\end{equation}
with $T_{c}=0.34$, $a_{0}=0.55$, $b_{0}=4.23$ and $\delta = 1.5$.
The solid line in Fig. 2(a) represents values calculated from this equation.
At $T=T_{c}$, $\psi(T)=0$ which means that for $T \leq T_{c}$ all bonded particles become $s$-particles.
The glass entropy $S_{g}(T)$ is calculated from the relation
\begin{equation}
S_{g}(T)=S_{g}(T_{1})-\int_{T}^{T_{1}} \frac{C_{vg}(T')}{T'} \mathrm{d}T' ,
\end{equation}
where $C_{vg}(T)=C_{v}(T)-\psi(T)^{2}$ and $S(T_{1}=1)$ is a constant chosen to raise the data. 
In Fig. 3 we plot $C_{v}$, $C_{vg}$, $S(T)$ and $S_{g}(T)$ vs 
$\frac{1}{T}$. Extrapolated values of $C_{v}$ and $C_{vg}$ shown by dotted lines are found, respectively, 
from an algebraic fit $C_{v}=1.5(T^{-2/5}+1)$ and subtracting from $C_{v}$ the extrapolated values of 
$\psi(T)^{2}$. As expected, $C_{v}$ and $C_{vg}$ meet at $T_{c}=0.34$ indicating that the
specific heat of the system below $T_{c}$ is entirely due to the glassy configurations.
The extrapolated values of $S(T)$ and $S_{g}(T)$ (see Fig. 3(b)) meet at $T\simeq 0.30$ which is close to the 
estimated value of the Kauzmann temperature $T_{k}\simeq 0.32$ \cite{Coluzzi}. The agreement shown in Fig. 3(b)
for $S_{g}(T)$ justifies our assumption about the role of the configurational fluctuations.

In Fig. 4 we plot $N_{\alpha}^{(b)}$, $N_{\alpha}^{(m)}$ and $N_{\alpha}^{(s)}$ vs $\frac{1}{T}$.
As temperature is lowered from $T=1.0$, all kind of particles initially increase, but close to $T\simeq 0.45$,
$N_{\alpha}^{(m)}$ has a tendency to decline while $N_{\alpha}^{(s)}$ to rise with increasing slope.
As indicated above, at $T=0.34$, $N_{\alpha}^{(s)}=N_{\alpha}^{(b)}$ and $N_{\alpha}^{(m)}=0$.
This indicates that there will be a very sharp rise in the value of $N_{\alpha}^{(s)}$ from $\sim 5$ at $T=0.45$
to $\sim 28$ at $T=0.34$. This sharp rise in $s$-particles and increase in values of barrier heights of
shells will sharply increase the activation energy and therefore the relaxation time. At present we do 
not have values of $g_{\alpha\gamma}(r)$ at lower temperatures to predict precise nature of the rise of 
$N_{\alpha}^{(s)}$ and $E_{\alpha}^{(s)}$ close to $T_{c}$. However, from above discussions it appears 
reasonable to suggest that temperature $T_{c}$ is what is known as the glass temperature $T_{g}$.

In conclusion, we demonstrated that a sharply increasing ``length'' which makes the relaxation time to rise 
sharply near the glass temperature can
be calculated from data of the radial distribution function. This ``length'' is the number of (non-chemical)
``stable bonds'' $N_{\alpha}^{(s)}$, a particle forms with its neighbors. The number $N_{\alpha}^{(s)}$
is found to depend on a part of the energy fluctuations referred to as the ``configurational fluctuations''
and denoted as $\psi(T) T$. The value of $\psi(T)$ can be determined from values of configurational entropy
which in turn can be calculated from inter-particle interactions \cite{Berthier}. In this work we, however,
used data of the relaxation time to calculate values of $\psi(T)$. From values of $\psi(T)$ we calculated the
configurational entropy at different temperature and compared with values determined from simulations 
\cite{Berthier}. The agreement between the calculated and simulated values validates our approach.
It is shown that close to the temperature $T_{c}$ where $\psi(T)$ goes to zero, $N_{\alpha}^{(s)}$ increases
sharply which in turn increases the relaxation time several order of magnitudes as happens near the glass 
temperature.
\\

We thank Smarajit Karmakar and Indrajit Tah for providing us the data of 
$g_{\alpha\gamma}(r)$ used in the present work.
The financial help from the Council of Scientific and Industrial 
Research and the Indian National Science Academy, New Delhi is acknowledged.

\end{document}